# The operational framework for quantum theories is both epistemologically and ontologically neutral

Laurie Letertre[a]

[a] Institut NEEL, 25 rue des Martyrs BP 166, 38042 Grenoble, France.

Email : laurie.letertre@neel.cnrs.fr

**Acknowledgments**

Many thanks to Vincent Lam and Cyril Branciard for their continuous encouragement and suggestions for the improvement of this paper. I am also thankful to Hippolyte Dourdent and Alexei Grinbaum for useful comments on a late version of this work. Finally, I am grateful to anonymous referees for helpful suggestions about the content of the paper. This work was supported by the Agence Nationale de la Recherche under the programme Investissements d'avenir (ANR-15-IDEX- 02).

# The operational framework for quantum theories is both epistemologically and ontologically neutral

**Abstract**

Operational frameworks are very useful to study the foundations of quantum mechanics, and are sometimes used to promote antirealist attitudes towards the theory. The aim of this paper is to review three arguments aiming at defending an antirealist reading of quantum physics based on various developments of standard quantum mechanics appealing to notions such as quantum information, non-causal correlations and indefinite causal orders. Those arguments will be discussed in order to show that they are not convincing. Instead, it is argued that there is conceptually no argument that could favour realist or antirealist attitudes towards quantum mechanics based solely on some features of some formalism. In particular, both realist and antirealist views are well accomodable within operational formulations of the theory. The reason for this is that the realist/antirealist debate is located at a purely epistemic level, which is not engaged by formal aspects of theories. As such, operational formulations of quantum mechanics are epistmologically and ontologically neutral. This discussion aims at clarifying the limits of the historical and methodological affinities between scientific antirealism and operational physics while engaging with recent discoveries in quantum foundations. It also aims at presenting various realist strategies to account for those developments.

**Keywords**: quantum mechanics, operational frameworks, realism, antirealism, indefinite causal order, process matrix formalism

## 1. Introduction

The debate between the two opposite stances that are scientific realism and antirealism is a central one in philosophy of science since the development of modern science (Psillos 2005). Scientific realism holds the view that our best scientific theories provide approximately true descriptions of the objective, external world. Such a philosophical attitude relies on three assumptions (Chakravartty 2017): (i) a metaphysical proposition according to which there exists a mind-independent, objective world, (ii) a semantic proposition according to which whether our theories are true or false is determined by the composition of nature and (iii) an epistemic proposition according to which science can provide access to objective knowledge of the external world. Denying any of those propositions leads to a form of antirealism about scientific theories. The most common form of scientific antirealism is epistemic, as it denies the epistemologically warranted and direct access to objective reality through science.

In the context of this paper, we will focus on the scientific realist and antirealist views as applied in the more specific context of quantum mechanics. A variety of antirealist approaches towards that theory have been developed. Those approaches range from a mere agnosticism regarding whether the ontology and dynamics postulated by the theory is approximately true (rejection of the epistemic proposition defined above), to a more radical stance claiming that quantum mechanics is not *about* objective reality. The underlying premises for the latter stance are either the rejection of the metaphysical and/or semantic propositions defined above, or the rejection of the epistemic proposition on the grounds that quantum mechanics is about a *subjective* reality.

The study of the foundations of quantum mechanics often appeals to operational formalisms. Those are frameworks in which any experiment can be represented independently of any physical model, and within which the theory can be recovered from a few basic physical principles. Such a framework is sometimes used to support an antirealist view towards quantum mechanics. I intend in this paper to defend the point that no formalism alone will ever provide evidence or support for either a realist or an antirealist stance, and that extra-assumptions are always needed in order to motivate a particular approach.

As such, this work is therefore not against antirealist readings of operational formalisms. Neither is it in favour of realist readings. Its goal and scope have a very specific focus: this discussion aims at providing a clarificatory emphasis on the fact that the way realist or antirealist approaches are philosophically motivated is no different in the context of operational formalisms than it is in the context of any other framework. Yet, I believe that these points deserve an explicit and clear formulation, not because they are necessarily polemical, but because they allow to highlight important non-explicit premises underlying recent antirealist approaches towards operational frameworks. Putting those under the spotlight will allow reclaiming a central fact: no new argument in favour of antirealism can be extracted from formal aspects in the foundations of quantum physics.

Through the discussion of particular antirealist arguments based on operational formalisms found in the literature, I will highlight the following points:

- The arguments attempting to support an antirealist view of quantum mechanics based solely on the operational formalism for physical theories alone are not convincing arguments (section 3).
- Similarly, there is no convincing argument for realist approaches to quantum mechanics grounded in the operational formalism in itself. Overall, both realist and antirealist approaches towards quantum mechanics are in principle equally compatible within the operational framework for physical theories (section 4).
- The reason for this *in principle* compatibility of both realist and antirealist views of quantum mechanics with operational formulations of the theory is that the realist/antirealist debate is purely epistemic and does not involve any feature of any formalism. The arguments involved are not tied to a specific theory (section 5).
- In particular, and this is the important point, the operational framework for physical theories is, in itself, epistemologically neutral, and, within a given epistemic framework, the interpretation of the theory (its ontological content) is postulated. The operational framework for physical theories *in itself*[1] does not favour a particular epistemic or ontological stance.

This discussion will relocate the realist/antirealist debate about quantum mechanics where it should be, i.e., at the epistemological level[2]. As a result, one cannot draw a realist or antirealist argument solely based on the features of some chosen formalism. The operational physics *in itself*, not only does not favour antirealist views over realist ones, but is also neutral epistemologically *and* ontologically.

Before exposing the above arguments, the next section will briefly present the operational framework for physical theories.

## 2. The operational framework for physical theories

---

[1] I.e. the purely formal content of the theory, without any additional interpretative considerations. A remark is worth mentioning here: it is true that the formal content of a theory can constrain any assigned ontology (e.g., the quantum ontologies designed within realist approaches to quantum mechanics are highly constrained by the formal aspects of the theory), and a particular ontology taken as a premise can influence the formal aspects of a theory in development (e.g. Bohmian mechanics was developed around a particular primitive ontology). Yet, such a mutual influence does not imply that a particular formalism needs to be interpreted in one particular way. There is not one unique ontology that can be postulated for any given formalism (see, e.g., a discussion in Chen (2019)). A given formalism, even designed with a specific interpretative framework in mind, can still be read along different lines by other thinkers. To sum up, even if *in the practice* of developing and interpreting a theory, there is a mutual dependence between the formal aspects and the postulated meaning of a theory, a given interpretation for a given formalism (whether it is a realist or an antirealist one) is never *necessary*.

[2] This point is actually a very general one, but this paper focus on quantum mechanics (and generalisations thereof) in particular because the philosophical implications related to those theories seem to be more pressing.

The *operational framework* for physical theories is also called the framework of *operational probability theories*[3] (OPT). In that framework[4], the *state* of a physical system is associated with the preparation procedure leading to that state. More precisely, a state is a *class* of operationally equivalent preparation procedures, i.e., procedures that cannot be distinguished experimentally. It is represented by a vector noted ω in a vector space. Any measurement procedure that can be performed on the system can be decomposed into a set of 1-bit measurements. Those are called *effects* and noted $e$[5]. They are represented by vectors in a second vector space (the "dual" space). A vector (state or effect) that cannot be expressed as a convex combination of other vectors is said to be *extremal*, or *pure*, while it is *mixed* otherwise. Performing a specific 1-bit measurement (corresponding to a specific effect) on a system in a given state yields the outcome 0 or 1, usually not in a reproducible way, but rather according to a certain probability distribution. Knowing that distribution for all the combinations of extremal states and effects is sufficient to calculate the outcome probability for any arbitrary pair of state and effect. For composite systems made of different single systems, an appropriate composition rule between the single system's state spaces on the one hand, and between the single system's effect spaces on the other hand, will determine how those sub-systems will interact with each other.

Hence, a physical experiment is entirely encapsulated in a probability distribution correlating the states of the system with the outcomes of the possible measurement's procedures. This probability distribution is defined by the sets of extremal states and effects and the correlations among each other. Upon postulating a new distribution, symmetries of the physical law to be reproduced, or some dynamical aspects of the system to be modelled might influence the resulting structure of the state and effect spaces. Importantly, those state and effect spaces have to satisfy consistency rules ensuring that the combination of any effect with any state yields a scalar compatible with a probabilistic interpretation, i.e., comprised between 0 and 1. Finally, the definition of the composition rule for composite systems will be crucial in determining whether the resulting probability distribution will display classical, quantum or post-quantum features, the latter being neither classical nor quantum.

A theory can be constructed by selecting an ensemble of probability distributions satisfying a set of basic axioms. Many toy theories have been constructed so far, displaying state spaces of various dimensions and shapes (Janotta et al. 2011, Janotta 2012, Janotta 2014 and the references therein). When a made-up theory allows to recover classical or quantum mechanics, it provides a reconstruction, also called *axiomatisation*, of that theory. The OPT framework is sufficiently general to include a large amount of possible probability theories, with classical and quantum theories as special cases.

Importantly, these basic axioms rely entirely on physical processes such as preparation and measurement procedures in an experimental setup. These principles are therefore based on *operational* processes, which explains why we talk of *operational* formulations of theories. Alternatively, some operational frameworks are also sometimes referred to as a *device-independent formalism*. Indeed, no specific machinery, tool or apparatus is mentioned in an operational probability theory: the experimental setup is reduced to a black box fed with some inputs and returning some output. No particular device is used within the theory. Finally, some of these basic operational axioms might be referred to as *information-theoretic* principles. Such principles involve processes and rules governing them that are at the core of quantum information theory, which studies how quantum systems can be used to process and communicate information (see Timpson (2013) for a review).

A pioneer axiomatisation of quantum mechanics from 5 axioms, some of which are inspired from quantum information theory, was made by Hardy (2001). Subsequent work provided further operational axiomatisations (Clifton et al. 2003, Chiribella et al. 2010, 2011, Dakic & Brukner 2011, Hardy 2011, Masanes & Muller 2011, Fivel 2012, Masanes et al. 2013, Barnum et al. 2014, Wilce 2019).

---

[3]  Also referred to as *generalised probabilistic theories*.

[4]  The present description of this framework is based on the work of Janotta & Hinrichsen (2014), Myrvold (2010), Timpson (2013), D'Ariano et al. (2017) and D'Ariano (2010).

[5]  More general transformations can also be considered in OPTs. Intermediate transformations of the system can be considered as being part of the preparation or measurement procedures.

## 3. Main antirealist arguments based on operational physics

This section will discuss three arguments to be found in the literature defending an antirealist view of quantum mechanics based on some features of the operational framework for physical theories. For each argument, a quick presentation is followed by objections.

While the second and third discussions are, to the best of my knowledge, new in the literature, the first argument's discussion is reviewed from previously existing work (Timpson 2013). It seems that this first argument remains somehow present in some informal conversations in the context of conferences in foundations of quantum mechanics, or in some work posterior to that of Timpson (see below). It seems therefore interesting to recall Timpson's objections in this paper. Moreover, including a review of Timpson's work in this discussion allows to show a richer variety of different strategies supporting antirealism towards quantum mechanics based solely on features of some formalism.
It will be interesting to see that all of these three arguments involve interpretational steps or assumptions that need to be considered in addition to the features of the operational formalism itself. These additional assumptions either boil down to more general considerations belonging to the realism/antirealism debate, or need further justifications to overcome the objections that are raised against them.

### 3.1 "Quantum mechanics is about 'quantum information', i.e., the wave- function is mere information"

#### 3.1.1 Summary of the argument

As announced above, the argument discussed in this section has been reviewed and objected to by Timpson (2013, Chap.7). It was recalled in section 2 that quantum mechanics can be reconstructed (axiomatised) from a limited amount of basic axioms that select the set of probability distributions satisfying each of these principles. This results in a particular operational probability theory that recovers standard quantum mechanics in the sense that the axioms select the distributions allowed by quantum mechanics, while excluding all the distributions that are not. In particular, quantum mechanics can be reconstructed (axiomatised) from informational principles, i.e., from axioms governing how quantum systems can be used to process and communicate quantum information. As expressed in the work of Timpson (2013, p. 150), such particular OPTs are sometimes taken as the sign that quantum mechanics is only "about information", hence, should be interpreted epistemically. In other words, quantum mechanics would not be directly about an external objective world, but about the *empirical knowledge, representation, or even beliefs* that we have of the world, which does not necessarily coincide with the actual objective world underlying those knowledge, representation or beliefs. This therefore amounts to a form of scientific antirealism as regards to quantum mechanics (we reject the epistemic proposition for scientific realism).

Specific examples of such a move can be found in (Zurek 1990, p. viii) or, more recently, in (Koberinski & Müller 2018). The latter acknowledges the interpretative neutrality of the operational (information-theoretic) formalism itself:

"Our arguments [...] support the hypothesis that quantum theory is a principle theory of information, with continuously-reversible evolution in time as a characteristic property. Any further insights into an underlying "quantum reality" (if it exists), or into the question "information about what" (if it has an answer) should not be expected to arise directly from these principles, or from quantum theory itself, but from a novel, yet-to- be-found constructive theory with additional beyond-quantum predictive power (if it exists)." (Koberinski & Müller 2018, p. 11)

However, it still mentions the reasoning described above:

"Interpretations that treat the quantum state as a state of knowledge or belief are conceptually more closely related to the view of quantum theory as a principle theory of information, which has led some physicists (e.g. Brukner) to argue that the success of the latter is evidence for the validity of the former." (Koberinski & Müller 2018, p. 5)

While it would not be fair to claim that they embrace fully and entirely such a view (as the first quote clearly shows), it is still interesting to see that this reasoning remains shared at least informally. Whether such an informal talk is meant to be taken literally or not, it remains a good illustration of the above-mentioned reasoning, which is problematic when taken at face value.

### 3.1.2 Objections

An objection to the above reasoning was articulated by Timpson (2013, Chap.7) as follows. First of all, the notion of "information" can take different meanings. We should differentiate a technical notion of information, which can be expressed in purely physical terms, from an "everyday sense" of information, associated with some elements of knowledge and language that are not expressible in purely physical terms. Then, Timpson makes clear that the notion of quantum information (appearing in information-theoretic reconstructions of quantum mechanics) is a "technical" notion of information. On the one hand, quantum information can be quantified using the quantum analogue of the classical Shannon information theory (see the work of Timpson (2013, Chap. 3) for a review). On the other hand, Timpson identifies a *piece* of quantum information as being merely a sequence of quantum states in a particular order. Hence, both the content of a particular piece and the quantification of quantum information can be expressed in purely physical terms, which shows the technical nature of quantum information. As a result, viewing quantum information as a kind of epistemic notion of information is, in itself, an unwarranted jump from a technical to an everyday notion of information. Such a jump needs further justification, which is not found in the formalism of operational physics alone.

We can expand a bit on this argument by Timpson. The formalism of a physical theory is, by itself, of technical nature. Interpreting the formalism is giving meaning to its technical elements. This implies to connect those elements to a wider conceptual context. More concretely, we can see that an epistemic view of the wavefunction refers to notions (such as observer, knowledge, belief, subjectivity, representation, ...) that need a language and background to be expressed and understood, that lies far beyond the scope of physics alone. Similarly, an ontic view towards the wavefunction involves concepts (such as objectivity, reality, laws, properties, ...) that cannot be captured by technical language within a physical theory alone.

To sum up, jumping from technical notions to epistemic ones (e.g. everyday notion of information) or ontological ones (e.g. objective entities) amounts to interpreting the notions under consideration. This jump requires philosophical justification that cannot be derived from any formalism alone. This point will be reiterated in section 5.

A final remark regarding this discussion concerns famous antirealist views built on, or developed along, information-theoretic formulations of quantum mechanics, namely the "it from bit" approach of Wheeler (Wheeler 1999) and Qbism (Fuchs et al. 2014)[6]. Those approaches do not need to rely on any semantic jump between different notions of information[7]. Yet, their association with information-theoretic formulations of quantum mechanics is very natural, which might suggest that the operational framework brings some evidence for an antirealist reading of quantum physics. However, as argued in (Timpson 2013) and (Felline 2018), it is very clear that Qbism, like any interpretation of quantum mechanics, is backed-up by a whole apparatus of philosophical premises that are not deduced from any formal content of quantum theory. In particular, Qbism is not deduced from operational physics, and its natural affinity with this formalism comes from its programme of relying on information-theoretic axiomatisations of quantum mechanics to identify the objective content of the theory (Fuchs 2004). Similarly, it is argued in (Timpson 2013, Chap. 3)[8] that the "it from bit" approach arguably needs specific philosophical background assumptions to be fully articulated and describe how our material world can be derived from a fundamental notion of quantum information. Those assumptions are not deduced from the operational formalism, which therefore needs to be supplemented with this additional philosophical machinery. Hence, the "it from bit" approach builds on information-theoretic formulations of quantum mechanics, but is not deduced from it, leaving room for alternative readings (see, e.g., section 4). With that being said, and as will be reminded in section 5, the convenience of the operational framework relative to a specific antirealist interpretation is not evidence for the likelihood of that interpretation, given that (i) equally convenient realist readings can be attributed to operational physics (see section 4), and (ii), even if the last point was not available, the empirical equivalence of a variety of formulations of quantum theory ensures that none has a privileged status when it comes to the interpretation of quantum mechanics.

---

[6] As it is well known, Qbism falls in the category of epistemic antirealist views. Indeed, it rejects the view that quantum mechanics provides an objective picture of the world as it is. While this does not imply a form of metaphysical antirealism or instrumentalism, standard quantum mechanics is said to provide an irreducibly subjective picture of reality (Fuchs & Schack 2014, Fuchs 2017). In the case of Wheeler's "it from bit" approach, it is often taken as an antirealist approach (see e.g. (Cabello 2017)). Standard quantum mechanics is seen as providing a "mere continuum idealization" (Wheeler 1999, p. 1). However, the precise way Wheeler's view is to be interpreted is debated, see e.g.(Fuchs & Schack 2014).

## 3.2 "Operational physics strongly suggests an epistemic interpretation for quantum probabilities"

### 3.2.1 Summary of the argument

The argument discussed in this section was *neutrally* reviewed[9] by Leifer (2014, section 2.3) and can be summarised as follows: there exists an operational probability theory which generalises the classical probability theory in such a way that it has both quantum theory and classical probability theory as special cases. Hence, in this particular OPT, the quantum probability distributions "play the same role" as the classical ones (Leifer 2014, p. 76). As a result, it seems natural to give them the same interpretation. Since classical probabilities are epistemic (as they originate from some ignorance of the observer regarding the exact physical state of the studied system), quantum probabilities should be seen as epistemic as well.

Does this argument support an antirealist view towards quantum mechanics? Interpreting the quantum probabilities as being of an epistemic nature means that the quantum state from which those probabilities are computed represents "a description of what an observer currently knows about a physical system. It is something that exists in the mind of the observer rather than in the external physical world. [...] The key property that this implies is that a given ontic state is deemed possible in more than one epistemic state. [...] [The quantum state] is simply a mathematical tool for determining probabilities, existing only in the minds and calculations of quantum theorists." (Leifer 2014, p. 69-71)[10]. While this antirealist view towards

---

[7] Within the Qbist interpretation, the quantum probabilities are given a Bayesian reading, and information-theoretic reconstructions of the theory are expected to give us knowledge about the *structure* of the theory, which in turn is hoped to inform us about some objective features of the world (Fuchs 2017). The motivations behind the Bayesian interpretation of quantum probabilities do not come from those information-theoretic reconstructions. Instead, the main motivations for Qbism find their root in the conceptual strength of the Bayesian interpretation of probabilities, as well as in the dissolution of the measurement problem and that of the issue of nonlocality (see (Timpson 2013, Chap. 9)).
   Wheeler's "it from bit" approach does not rely on a semantic jump between the technical and everyday senses of information since, within that view, quantum information is assimilated to none of these notions. Instead, quantum information has an ontological status as a fundamental entity from which every physical entity derives its existence (Wheeler 1999).

[8] Felline (2018, p. 5) also provides an assessment of the philosophical premises behind an ontic reading of informational-theoretic formulations of quantum mechanics (which is a (weak) form of realism towards quantum mechanics in which the theory is about some new objective stuff that is quantum information), that might be relevant to the "it from bit" approach.

[9] I do not claim anything about the actual stance of Leifer regarding that argument.

[10] This epistemic notion of the wavefunction found in (Leifer 2014, p. 69-71) is referred to as the "$\psi$-epistemic" reading of the wavefunction, as opposed to a "$\psi$-ontic" reading, in which the wavefunction refers to something objective in the world (Leifer 2014). To avoid

the *quantum state* does not necessarily entail an antirealist view of quantum mechanics itself, it rules out the main realist accounts of quantum mechanics (namely Bohmian mechanics, the GRW theory and Everettian mechanics) as those are committed to an ontic conception of the wavefunction (Cabello 2017), which is then considered to refer to objective elements of reality[11].

### 3.2.2 Objection

A given *interpretation* of probabilities in the context of a given theory (in our case a particular OPT) is about the **meaning** of the probabilities (e.g. some agent's ignorance (epistemic origin) or the existence of some fundamental randomness in nature (ontic origin)), and this meaning depends (among other things) on the kind of objects and dynamics that are involved in the theory.

Yet, although what is meant by "quantum mechanics and classical probability theory are special cases of one more general OPT" is that we can use the same mathematical tools to express both classical and quantum theories, those mathematical tools do not necessarily carry the same ontological meaning in each case. In this particular OPT, a set of basic operational axioms allows for a wide range of correlations. Those correlations involve particular objects with particular dynamics, and all display the same operational features, namely those formulated in the axioms. These operational characteristics happen to encompass both classical and quantum correlations. Yet, there is no necessity for the objects and dynamics involved in the correlations to be of the same nature in the classical and quantum cases[12]. A straightforward example is the spin of an electron, a property possibly involved in a quantum correlation that has no classical counterpart.

Another reason to doubt that an OPT encompassing the quantum and classical theories describes objects and dynamics of the same nature is that it can be evidenced within the operational framework that there exists a fundamental distinction between quantum and classical operational theories, due to the presence of entanglement. In the review of Janotta & Hinrichsen (2014), it is explained how any non-classical probability distribution for a single system formulated in the framework of OPT can be expressed as a classical distribution by increasing the dimension of the state space. Conversely, any classical single system's distribution can transform into a distribution with non-classical features by imposing additional restrictions on its effect vectors, which lowers the dimensions of its state space. This result questions the fundamental necessity of the classical/quantum distinction, since non-classical theories seem to be always reducible to classical ones by resorting to higher-dimensional vector spaces. However, Janotta & Hinrichsen (2014) also explain that this situation is no longer possible for composite systems' distributions, which cannot be expressed classically, no matter the dimensionality of its vector spaces. We can therefore conclude that there exist genuinely non-classical systems, fundamentally distinct from classical ones. The existence of quantum entanglement can possibly mean that there are different dynamics at play between the classical and quantum realms. Hence, quantum mechanisms among quantum objects different from the classical ones might be taking place, which would mean a possibly different origin (and therefore interpretation) for the measured probabilities.

---

any confusion, it is worth mentioning that those terms "$\psi$-epistemic" and "$\psi$-ontic" have been used and defined quite differently by some authors posteriorly to Leifer's work, e.g. in (Cabello 2017), where "$\psi$-ontic" has the same meaning as in (Leifer 2014, p. 69-71) but "$\psi$-epistemic" is a more restrictive term that is actually applicable to the realist views in which the wavefunction is "representing knowledge about an underlying objective reality". In the present context, it is Leifer's broader notion of the $\psi$-epistemic view that is considered. As such, Leifer's notions of $\psi$-epistemic and $\psi$-ontic views are independent of the distinction between realist and antirealist approaches to quantum mechanics. Indeed, both types of views can be found in either of realist or antirealist approaches to the theory.

[11] The question of whether the wavefunction is indeed ontic (i.e. refers to something objective in the world (Leifer 2014)) within the main realist accounts of quantum mechanics arguably depends on the particular ontology given to the theory. Indeed, if one subscribes to a nomological reading of the wavefunction, while not assigning any objectivity to laws of nature, strictly speaking it cannot be said that the wavefunction has an ontic status (Gao 2019). However, this scenario is a very specific one among a variety of existing accounts.

[12] It may seem here that the argument presupposes realism. However, this reasoning holds whether the meaning we assign to the theory is of a realist or antirealist nature. Whether the theory is considered to be about objective entities or not, we need to specifiy what it refers to in order to take a stance about the meaning of its probabilities. The notion of "object" considered here is taken very broadly, encompassing any posits that could correspond to a realist or an antirealist view.

In summary, since the OPTs encompassing both quantum and classical theories do not necessarily involve a unique kind of objects and dynamics, then OPTs do not necessarily involve a unique interpretation for the probability distributions. This point is overlooked in the antirealist argument because of the implicit character of operational physics regarding objects, properties and underlying dynamical laws.

As a concluding remark, we can quote the statement of Timpson (2013, p. 187) regarding the ontological significance of operational theories:

"The form of a theory convenient for placing it within a space of theories need not be the fundamental, ontologically revealing, form of the theory: identifying properties of the theory (the former setting) and saying how the world is (the latter) are different tasks; and it is not at all surprising that we might expect different formal representations of the theory to be more or less useful for these distinct tasks. An operationalist black-box formulation of a theory might be most appropriate for the former task, but there is no reason at all why that should be taken to be the final story, the end of discussion about the theory: there is still an ontological story to be told too—the underlying dynamics giving rise to the results schematised within the story of black boxes."

The operational formulation of quantum mechanics is an alternative way of modelling the same physics than the standard (model-dependent) formulation, emphasising correlations between inputs and outputs of experiments while the underlying mechanisms involving objects and their dynamics is left unspecified. Yet, it is this underlying story that is of interest when wondering about the meaning of quantum probabilities. The existence of an OPT encompassing both classical and quantum correlations merely means that one can find a set of common axioms (operational properties) satisfied by these correlations. It therefore gives information about the similarities that exist between the *structures* of the two theories. However, it does not entail that the same mechanisms underlay both classical and quantum correlations. As a result, it does not imply a same ontological story for classical and quantum theories.

### 3.3 "Device-independent physics evacuates the notion of systems, therefore physics is not about systems"

#### 3.3.1 Summary of the argument

The argument that will be discussed is found in (Grinbaum 2017), in which two claims are made:

- *Claim 1*: "Incompatible with the old explanatory mode, device-independent[13] models typically do not meet the conditions for the emergence of robust theoretical constituents corresponding to real objects. By allowing no room for systems, they inaugurate the obsolescence of this elementary building block [...]. " (Grinbaum 2017, p. 3)

- *Claim 2*: "[...] physical theory is about languages: it is defined by a choice of alphabets for the inputs and the outputs [of a given experiment] and by the conditions imposed on this algebraic structure. Strings, or words in such alphabets, form a common mathematical background of device-independent approaches. [...] If strings are not 'about' some elements of reality, they can be said to be 'about' languages from which they are formed." (Grinbaum 2017, p. 14 & 16)

The first claim is defended by appealing to two further statements: the notion of systems in a particular operational formulation of quantum physics introduces difficulties (statement A). Yet, the notion of physical systems is not a necessary ingredient in the process of interpreting a theory (statement B). Since it is auxiliary *and* problematic, we should get rid of it.

---

[13] A sub-category of operational formalisms that do not appeal to any particular device to represent the described experiments.

Statement B is stated in Grinbaum's work on the grounds that what is really necessary to describe an experiment are the notions of input and output linked to a given party (which can be seen (in spite of their different philosophical flavour) as the counterparts of the notions of preparation procedure, measurement result and experimental setup seen in section 2, respectively), while spatiotemporally defined notions such as physical systems are mere interpretative devices (Grinbaum 2017, p. 7-8).

Statement A is, for its part, defended as follows: Grinbaum refers to a particular operational theory, called process matrix formalism (Oreshkov et al. 2012), which generalises quantum mechanics by dropping the assumption of a global causal structure relating the inputs and outputs among different parties. In other words, while there is a local temporal ordering in a given party allowing to claim that some output temporally succeeds some input of that party, we make no claim regarding the ordering of inputs and outputs pertaining to different parties. Operationally, a theory for which no global causal order among parties is assumed may allow for the prediction of joint probability distributions for which there is no *definite* global causal order. We speak in that case of *indefinite causal order*, i.e that this distribution is *incompatible* with any definite causal ordering among parties, as expressed within the formalism[14]. Grinbaum claims that the notion of physical systems runs into three difficulties in the context of this particular theory (Grinbaum 2017, section 3):

- The absence of global causal structure makes it impossible to make sense of the spatiotemporally defined notion of physical system (as being spatially delimited at all times from the rest of the environment) that would endure and evolve across the multipartite experiment.
- The absence of global causal structure allows for strange situations where "'systems' in the process matrix framework may 'enter' the same local laboratory twice", which Grinbaum considers to be a situation that never happens to a physical object.
- A framework allowing for an absence of global causal structure predicts indefinite causal orders not solely for some quantum correlations, but also for certain multipartite "classical"[15] correlations, which shows that the difficulties faced by the notion of physical systems do not arise exclusively in a quantum context.

Grinbaum concludes from these three points that the notion of physical system is problematic in the case of indefinite causal orders.

Now, regarding the second claim of the main argument, Grinbaum postulates that physics is about languages, on the grounds that such a proposition provides a common philosophical background for *all* physical theories, and that this background ontology is minimal, yet sufficient to answer the question "What are physical theories about?". Such a view can be read as a form of scientific antirealism, since according to that position, science ceases to speak about an objective external reality independently of the language used to describe it. Instead, science is about language itself. In particular, Grinbaum explicitly mentions that the kind of language that he considers is of a formal nature, rather than of the everyday kind (Grinbaum 2017, p.16).

### 3.3.2 Objections

The point is not to criticise the ontology proposed by Grinbaum (claim 2), nor its overall project (Grinbaum 2007), but to object to a very specific part of his argument, namely the idea that considerations from the operational formalism *alone* preclude realist accounts of quantum physics appealing to physical systems (claim 1).

---

[14] As will be discussed later, the question of whether causal relations are indeed definite or not (ontologically speaking) is not straightforward. The answer will depend on the particular reading (interpretation) that will be given to the theory.

[15] A word of caution is needed regarding the use of the term "classical". Some non-causal correlations can indeed be said to be "classical" in the sense that their corresponding density matrices are diagonal, which means that the probabilities can be interpreted classically. However, in a more stringent way, these correlations are undoubtedly nonclassical. Indeed, any non-causal correlation lies outside the polytope of classical correlations, for which everything is well-defined.

In the previous section, we recalled Grinbaum's list of the various difficulties that the notion of system introduces when we deal with the notion of indefinite causal order. Yet, the core problem faced by the notion of physical system in the context of indefinite causal orders is stated in the first element of that list, and this difficulty points towards the fact that a realist account of quantum physics necessarily needs to revise our initial conception of how causal correlations are to be understood/conceived.

However, to this day, this challenge has not proved to be insurmountable, and future work will tell us what are the realist implications about quantum causality. Similar challenges are already present in standard quantum mechanics, in which a realist attitude towards nonlocality forces us to choose among a range of deep metaphysical implications such as retrocausation, the existence of a preferred frame in the universe or strange fundamental ontologies for the world (Maudlin 2011). Such metaphysical questions only become more pressing in more general operational theories, which predict new phenomena such as indefinite causal orders.

In that new context, any realist account needs to undertake the following question: are indefinite causal orders objective, or are they purely theoretical notions with no objective counterpart? So far, indefinite causal orders have not been observed purely operationally in terms of physical correlations violating the causal equivalent of a Bell inequality (Oreshkov et al. 2012, Branciard et al. 2015). For the purpose of his argument, Grinbaum takes their existence as a presupposition. Yet, to this day, their objective existence or non-existence remain to be proven.

If we do take indefinite causal order to exist in nature, the very question of whether spatiotemporal notions indeed become ontologically indefinite in such a situation remains open. Grinbaum's claim about the absence of spacetime background (regarding to which enduring and evolving spatiotemporally localised objects can be defined) is actually not a necessity, but an interpretational possibility. As a counter-example to this scenario, it seems possible to make sense of spatiotemporally localised objects in the context of indefinite causal orders by appealing to some forms of holism of the dynamical properties of physical objects.

Alternatively, one could also investigate the possibility to conceive the notion of physical system in a less stringent way than what is considered by Grinbaum (which promotes the idea of a spatially well-delimited object that endures through time). There exist ontologies that do not rely on objects enduring through time (such as the flash ontology developed in the context of the GRW theory (Ghirardi 2018, Allori et al. 2008)), or that do not have an intrinsic identity (such as some ontologies involving a relational notion of objects as in some ontic structural realist views (Ladyman 2020, section 4)). Those particular ontologies of objects could possibly make sense of a notion of physical system that would not be problematic in the context of an indefinite causal order.

Finally, even in the scenario of a non-fundamental spacetime background (which is actually a possibility that is seriously contemplated in certain research programmes in quantum gravity (Huggett & Wüthrich 2013)), this would not necessarily mean that the notion of physical system should be given up altogether (in the sense described by Grinbaum). Indeed, there are different ways to articulate fundamental ontologies not based on spatiotemporal notions, from which our familiar notion of physical system would emerge (Lam & Wüthrich 2018, Huggett & Wüthrich 2013, Le Bihan 2018). More precisely, as argued in (Lam & Wüthrich 2018), the most pressing challenge to meet when faced with a theory in which spatiotemporal concepts would be emergent (instead of fundamental), is to guarantee a formal and philosophical connection between the fundamental non-spatiotemporal entities postulated by the theory and the empirical realm (which seems to unavoidably involve spatiotemporal notions), so that the theory can actually be tested. In other words, it requires a *formal* consistency between the theory and that of general relativity (our current best theory of spacetime), but it also requires a *metaphysical* account of how spatiotemporal notions can emerge from non-spatiotemporal ones. This twofold requirement can be referred to as a requirement of empirical coherence (Huggett & Wüthrich 2013). Multiple strategies have been offered to reach (at least part of) that objective. Huggett & Wüthrich (2013) have argued that, given a formal theory in which spacetime would be emergent (and provided that the theory is true, which presupposes scientific realism), a formal derivation of the spatio-temporal notions from the non-spatiotemporal ones would secure in itself theoretical and conceptual empirical coherence. Still, spelling out this emergence metaphysically remains a challenge[16]. To undertake this task, as examples, a compositional (mereological) and functional approaches have been proposed by Le Bihan (2018) and Lam & Wüthrich (2018), respectively.

In conclusion, the new notion of indefinite causal orders allowed by a particular operational probability theory generalising quantum mechanics does not force us to abandon scientific realist accounts of quantum physics. Grinbaum's argument is a proposal among other possible ones, rather than a conclusive argument ruling out specific realist accounts of quantum physics.

Overall, the three arguments discussed above for an antirealist view of quantum mechanics based on the operational framework for physical theories alone are not conclusive arguments. At best, they are interpretative possibilities, or "antirealist readings" of the operational formalism, which is not "already containing" hints for such an antirealist view. We should acknowledge the specific philosophical assumptions that are presupposed for such anti-realist readings of operational quantum mechanics.

## 4. Realist approaches in the context of operational quantum mechanics

We saw in section 3 that the meaning of the probabilities in the operational framework are left "untouched/uninterpreted", and for that reason, the various arguments for antirealist views are not convincing when grounded in the operational formalism alone. Indeed, it was emphasised in section 3.1 that the notion of quantum information used in information-theoretic approaches is of a technical nature, and extra assumptions need to be added if one wants to give it an antirealist flavour. It was defended in section 3.2 that the meaning of probabilities was tied to a certain extent to the theory's ontology: whether we postulate that the theory is about objective entities or not, and whether those entities are all of the same nature with the same dynamics and properties or not, it is still an extra layer that we apply *on top* of the bare formalism. Finally, section 3.3 highlighted the fact that, in order to make sense of the formal notion of indefinite causal order, we need to postulate a meaning for the theory, whether realist or antirealist. It was argued that there is no necessity to follow either of those two attitudes. In summary, extra-assumptions were needed in all these three cases to jump from the operational formalism to an antirealist reading.

This section will now emphasise that the situation is similar for realist approaches. Indeed, any realist reading of the notion of quantum information needs an extra-assumption containing a postulated ontology for the theory under consideration. Interpreting the meaning of probabilities in a specific OPT requires to postulate an ontology (extra-assumption) for the objects involved in the described correlations. Explaining the notion of indefinite causal order in a realist fashion implies to postulate an appropriate ontology as well. Yet, a bare formalism does not impose a particular ontology, *a fortiori* a realist one. To sum up, no matter the approach (realist or antirealist), we must postulate a meaning which is not prescribed by the formal characteristics of the theory.

Hence, for similar reasons to the ones discussed in section 3, there is no convincing argument favouring realist approaches to quantum mechanics grounded in the operational formalism in itself. Yet, such realist approaches are not only possible ((Grinbaum 2007), for that matter, recommends that realist approaches take certain axiomatic reconstructions as a starting point), but possibly well accommodated by that formalism. As discussed in section 3.2, the operational framework puts the emphasis on correlations. While some antirealist views interpret these correlations as describing either *possible* objective (yet unwarranted epistemologically) phenomena affecting real entities, or irreducibly subjective notions based on the observer's experience of reality, realist approaches can view these correlations as describing real objective (epistemologically warranted) phenomena among different objects. Various interpretations specify different kinds of underlying mechanisms explaining those phenomena.

---

[16] This point is discussed in (Ney 2015) in the context of wavefunction realism, where the standard 3-dimensional spacetime emerges from the high-dimensional space in which the wavefunction lives.

As an example, a proposition in the realist camp, called ontic structural realism (Ladyman 1998, Lam 2017, Ladyman 2020, and references therein), claims a particular status for structures (as a set of physical relations) in the fundamental ontology of the world. Such a proposal applied to quantum mechanics, while not constituting a complete interpretation in itself[17], can be seen as an interpretative tool to articulate the metaphysical implications of quantum mechanics in a coherent way. Because the operational framework puts the emphasis on correlations, hence on physical relations connecting different systems, ontic structural realism seems to be a particularly well suited/convenient way to interpret the operational framework for physical theories. Such a strategy was also considered in (Koberinski & Müller 2018).

In conclusion, each realist/antirealist camp can **read** the operational formulartion of quantum mechanics according to their own view. The operational formalism in itself, not only does not explicitly favour a particular view, but also does not present any particular difficulty for any of them either. That point is further discussed in the last section of this paper.

## 5. The realist/antirealist debate is formalism-independent

We have seen in previous sections that extra-assumptions were needed to go from a bare formalism to an interpretation of the theory. These extra-assumptions provide a meaning to the theory, which agrees with either a realist or an antirealist view. Both possibilities are equally well accommodated by operational formalisms because those extra-assumptions are motivated by purely philosophical considerations, that are postulated on top of sole considerations from specific formulations of theories.

The idea that arguments in favour of epistemic antirealism are formalism-independent is a very old and uncontroversial view: they are based on general considerations about the dynamics and methods of science and experimental success of scientific theories, and are independent from the particular *form* or *content* of a theory. Antirealism about quantum mechanics will discard realist proposals postulating a direct link between the theory's ontology and that of the objective world on the grounds of more general considerations such as the "pessimistic meta-induction" argument (Chakravartty 2017, section 3.3), or on the grounds that their implications for reality are unavoidably implausible or too underdetermined (Van Fraassen 1980), or on the grounds that an antirealist approach is ontologically/metaphysically more cautious (Grinbaum 2009). The formalism-independent nature of the argument is obvious in the case of the pessimistic induction argument, and in the case of the underdetermination of interpretations of quantum mechanics. The argument favouring antirealist approaches on the grounds that they do not need to postulate any ontological content for the world is again an argument of an epistemic nature (Grinbaum 2009). The fact that this argument pairs up well with operational formulations of quantum mechanics that do not appeal to any particular model for the described experiments does not change its epistemic status.

Yet, it is interesting to make this point explicit in the case of the antirealist argument based on various kinds of (what are sometimes called) 'no-go theorems'. This argument, as it is well-known, contemplates the implications for the nature of reality when adopting a realist approach in the light of various theorems such as Bell's theorem Bell (1964), the Kochen–Specker theorem (see Kunjwal & Spekkens (2015) for the operational version), or more recently Shrapnel-Costa's theorem (Shrapnel & Costa (2018)). The implications of those theorems for realist models are considered to be too implausible by some antirealists, and form "no-go results" that should be taken as a hint that the theory is to be interpreted on different grounds, namely by rejecting the idea that the theory is about an objective reality (see Wiseman &

---

[17] An ontic structural realist reading of quantum mechanics does not provide alone a solution to the measurement problem, and can be used in the context of different realist interpretations of the theory (Lam 2017).

Cavalcanti (2017) for a discussion). Because those theorems are/can be expressed within the operational framework, it is useful to illustrate explicitly how such an antirealist argument relies on assumptions independent of the formalism itself.

Let's reconsider the case of Bell's theorem, while the narrative is the same for the other theorems. The field of application of this theorem is broader than quantum mechanics, and it applies to *abstract models* satisfying certain kinds of criteria. As famously known, in a nutshell, Bell's theorem states the following: "local models" imply certain empirically testable predictions (Bell 1964). Experiments show that those predictions are violated. Therefore, there is no local model that can account for experience. As a result, successful realist models of quantum mechanics are necessarily nonlocal. The demonstration of Bell's theorem does not appeal to the particular formalism of the models, what this formalism represents and what kind of dynamics is described. The reasoning only appeals to an abstract notion of models, i.e. an unspecified formalism that generates probability distributions as predictions. Among the assumptions required to demonstrate the theorem, the ones constraining the models themselves are their empirical success, their use of a variable λ that describes entirely the observed system, and the requirement of local causality. In that context, it does not matter whether we speak of the standard ("Hilbert space") formulation of quantum mechanics, or of a particular operational axiomatisation of the theory. It is mainly experimental considerations (modulo the extra-assumption often referred to as the "free choice" of the measurement settings[18]) that imply nonlocal features of theoretical models accounting for quantum experiments. Realists found various ways to develop a world's ontology in which nonlocality is an objective feature of nature (e.g. quantum ontologies of Bohmian mechanics, GRW theory or Everettian mechanics, ...). Those who claim that the idea of nonlocality as an objective feature of nature is not plausible will reject any realist model, irrespectively of the particular formalism used to express that model.

This reasoning holds as well for the Kochen–Specker (in its operational form) and Shrapnel- Costa's theorems. The operational formulation of the Kochen–Specker theorem (and the more general version of Shrapnel and Costa) shows that any model empirically successful is necessarily contextual. Their results therefore set particular constrains on any ontological model that would agree with the predictions of quantum physics. To reach those conclusions, they rely on a conjunction of general physical principles and empirical predictions.  Formally speaking, they appeal to abstract models, which makes them independent from any formalism. These results can either be taken as information about the objective world or not. Such a realist or antirealist stance is motivated by purely epistemic considerations.

To conclude, in the same way that antirealist arguments based on no-go theorems expressed in operational frameworks crucially appeal to certain premises that are not derived from (nor constrained by) the *formal* features of the formalism (namely the unplausible character of objective nonlocal features in nature), the antirealist arguments reviewed in this paper appeal to hidden extra-assumptions that are formalism-independent, whether they fall under the scope of metaphysics, semantics or epistemology. Operational frameworks are epistemologically and ontologically neutral in themselves.

## 6. Conclusion

This paper briefly presented the operational framework for physical theories and reviewed three arguments according to which operational formulations of quantum mechanics contain hints supporting an antirealist reading of the theory. Objections were provided, allowing to conclude that those arguments were not convincing. It was also argued that the operational framework was not providing any arguments for favouring a realist reading over antirealist ones. It was recalled that such results are expected, given the fact that the scientific realist/antirealist debate in the context of quantum mechanics is located at an epistemic level, and is not concerned by the specific form and content of the theory.

---

[18] This assumption is usually taken for granted, but has been critically discussed in the literature, see e.g. a review of past discussions in (Berkovitz 2016, and references therein), or more recent discussions in (Hall 2010, Friedman *et al.* 2019).

This whole discussion leads us to our main claim, namely that the operational framework for physical theories is both epistemologically and ontologically neutral in itself. First, the operational framework is epistemologically neutral since the arguments in defence of an epistemological stance towards quantum physics do not appeal to any formalism in particular; their success is not reinforced or lessened in the operational formulation of quantum mechanics compared to the situation in the standard formulation. Second, the operational framework alone is ontologically neutral since going from operational postulates to a proposal about the theory's ontology implies specifying the status of the correlations at the centre of the formalism, this status being postulated on top of the formal aspects of the theory.

Yet, this does not mean that the operational framework cannot be informative in some sense. While it will not allow us to discriminate between philosophical interpretations themselves, this framework can bring new knowledge. Formally, it allows us to learn about the structures and foundations of theories. Philosophically, the epistemic analysis of such formal information and the way it can be put to use within realist and antirealist approaches is to be pursued in future work.

## 7. Acknowledgments